\documentclass[preprint,showkeys,showpacs,preprintnumbers,amsmath,amssymb,tightenlines]{revtex4}
\usepackage{amsfonts}
\usepackage{graphicx}
\usepackage{natbib}
\usepackage{dcolumn}
\usepackage{bm}

\newtheorem{lemma}{Lemma}
\newenvironment{proof}[1][\noindent Proof]{\textbf{#1.} }{\ \rule{0.5em}{0.5em}}

\begin{document}

\preprint{}

\title{Directed or Undirected? A New Index to Check\\ for Directionality of Relations in Socio-Economic Networks\\}

\author{Giorgio Fagiolo}
 \altaffiliation{Sant'Anna School of Advanced Studies, Pisa, Italy.
 Mail address: Sant'Anna School of Advanced Studies, Piazza Martiri della Libert\`{a} 33, I-56127 Pisa, Italy.
 Tel: +39-050-883343 Fax: +39-050-883344}
 \email{giorgio.fagiolo@sssup.it}

\date{March 2007}

\begin{abstract}
\noindent This paper proposes a simple procedure to decide whether
the empirically-observed adjacency or weights matrix, which
characterizes the graph underlying a socio-economic network, is
sufficiently symmetric (respectively, asymmetric) to justify an
undirected  (respectively, directed) network analysis. We
introduce a new index that satisfies two main properties. First,
it can be applied to both binary or weighted graphs. Second, once
suitably standardized, it distributes as a standard normal over
all possible adjacency/weights matrices. To test the index in
practice, we present an application that employs a set of
well-known empirically-observed social and economic networks.
\end{abstract}

\keywords{Social Networks, Complex Networks, Directed vs.
Undirected Links, Symmetric Matrices, Statistical Properties.}

\pacs{89.75.-k, 89.65.Gh, 87.23.Ge, 05.70.Ln, 05.40.-a}

\maketitle

\section{Introduction\label{Section:Introduction}}

\noindent In the last years, the literature on networks has been
characterized by exponential growth. Empirical and theoretical
contributions in very diverse fields such as physics, sociology,
economics, etc. have increasingly highlighted the pervasiveness of
networked structures. Examples range from WWW, the Internet,
airline connections, scientific collaborations and citations,
trade and labor market contacts, friendship and other social
relationships, business relations and R\&S partnerships, all the
way through cellular, ecological and neural networks
\cite{AlbertBarabasi2002,Newman2003,WassermanFaust1994,Carrington2005}.

The empirical research has thoroughly studied the (often complex)
topological properties of such networks, whereas a large number of
theoretical models has been proposed in order to investigate how
networks evolve through time \cite{DoroMendes2003}. Structural
properties of networks have been shown to heavily impact on the
dynamics of the socio-economic systems that they embed
\cite{Watts1999}. As a result, their understanding has become
crucial also as far as policy implications are concerned
\cite{Grano1974}.

The simplest mathematical description of a network is in terms of
a graph, that is a list of nodes $\{1,2,...,N\}$ and a set of
arrows (links), possibly connecting any two nodes
\cite{Hara1969,Bollo1985}. Alternatively, one can characterize a
network through a $N\times N$ real-valued matrix $W=\{w_{ij}\}$,
where any out-of-diagonal entry $w_{ij}$ is non-zero if and only
if an arrow from node $i$ to $j$ exists in the network. Entries on
the main diagonal are typically assumed to be all different from
zero (if self-interactions are allowed) or all equal to zero (if
they are not). Networks are distinguished in binary (dichotomous)
or weighted. In binary networks all links carry the same
intensity. This means that in binary networks a link is either
present or not, i.e. $w_{ij}\in \{0,1\}$. In this case, $W$ is
called an ``adjacency'' matrix. Weighted networks allow one
instead to associate a weight (i.e. a positive real number) to
each link, typically proportional to its interaction strength or
the flux intensity it carries
\cite{Barr04,Barr05,Bart05,DeMontis2005}. Any non-zero entry
$w_{ij}$ thus measures the weight of the link originating from $i$
and ending up in $j$, and the resulting matrix $W$ is called the
``weights'' matrix \footnote{In what follows, we will stick to the
case $w_{ij}\in [0,1]$, all $i,j$ (the more general case
$w_{ij}\in R_{+}$ can be reduced to the former simply by dividing
all weights by their maximum level in $W$).}.

Both binary and weighted networks can be undirected or directed.
Formally, a network is undirected if all links are bilateral, i.e.
$w_{ij}w_{ji}>0$ for all $i\neq j$. This means that in undirected
networks all pairs of connected nodes mutually affect each other.
One can thus replace arrows with non-directed edges (or arcs)
connecting any two nodes and forget about the implicit directions.
This greatly simplifies the analysis, as the tools for studying
undirected networks are much better developed and understood.
Directed networks are instead not symmetric, as there exists at
least a pair of connected nodes wherein one directed link is not
reciprocated, i.e. $\exists (i,j), i\neq j: w_{ij}>0$, but
$w_{ji}=0$. Studying the topological properties of directed
networks, especially in the weighted case, can become more
difficult, as one has to distinguish inward from outward links in
computing synthetic indices such as node and average
nearest-neighbor degree and strength, clustering coefficient,
etc.. Therefore, it is not surprising that the properties of such
indices are much less explored in the literature.

From a theoretic perspective, it is easy to distinguish undirected
from directed networks: the network is undirected if and only if
the matrix $W$ is symmetric. When it comes to the empirics,
however, researchers often face the following problem. If the
empirical network concerns an intrinsically mutual social or
economic relationship (e.g. friendship, marriage, business
partnerships, etc.) then $W$, as estimated by the data collected,
is straightforwardly symmetric and only tools designed for
undirected network analysis are to be employed. More generally,
however, one deals with notionally non-mutual relationships,
possibly entailing directed networks. In that case, data usually
allow to build a matrix $W$ that, especially in the weighted case,
is hardly found to be symmetric. Strictly speaking, one should
treat all such networks as directed. This often implies a more
complicated and convoluted analysis and, frequently, less
clear-cut results. The alternative, typically employed by
practitioners in the field, is to compute the ratio of the number
of directed (bilateral) links actually present in the networks to
the maximum number of possible directed links (i.e. $N(N-1)$). If
this ratio is ``reasonably'' large, then one can symmetrize the
network (i.e. making it undirected, see
\cite{WassermanFaust1994,Pajek2005}) and apply the relevant
undirected network toolbox.

However, as shown in Ref. \cite{GarlaSymm2004}, this procedure has
several drawbacks. In particular, it is heavily dependent on the
density of the network under analysis (i.e., the ratio between the
total number of existing links to $N(N-1)$).

Moreover, and most important here, if the network is weighted, the
ratio of bilateral links does not take into account the effect of
link weights. Indeed, a bilateral link exists between $i$ and $j$
if and only if $w_{ij}w_{ji}>0$, i.e. irrespective of the actual
size of the two weights. Of course, as far as symmetry of $W$ is
concerned, the sub-case where $w_{ij}>>0, w_{ji}\simeq 0$ will be
very different from the sub-case where $w_{ij}\simeq w_{ji}>0$.

In this paper, we present a simple procedure that tries to
overcome this problem. More specifically, we develop a simple
index that can help in deciding when the empirically-observed $W$
is sufficiently symmetric to justify an undirected network
analysis. Our index has two main properties. First, it can be
applied with minor modifications to both binary and weighted
networks. Second, the standardized version of the index
distributes as a standard normal (over all possible matrices $W$).
Therefore, after having set a threshold $x$, one might conclude
that the network is to be treated as if it is undirected if the
index computed on $W$ is lower than $x$.

Of course, the procedure that we propose in the paper is by no
means a statistical test for the null hypothesis that $W$ involves
some kind of symmetry. Indeed, one has almost always to rely on a
single observation for $W$ (more on that in Section
\ref{Section:Conclusions}). Nevertheless, we believe that the
index studied here could possibly provide a simple way to ground
the ``directed vs. undirected'' decision on more solid bases.

The paper is organized as follows. In Section
\ref{Section:Definition} we define the index and we derive its
basic properties. Section \ref{Section:Statistical_Properties}
discusses its statistical properties, while in Section
\ref{Section:Examples} we apply the procedure to the empirical
networks extensively studied in \cite{WassermanFaust1994}.
Finally, Section \ref{Section:Conclusions} concludes.

\section{Definition and Basic Properties\label{Section:Definition}}

\noindent Consider a directed, weighted, graph
$\widetilde{G}=(N,\widetilde{A})$, where $N$ is the number of
nodes and $A=\{\widetilde{a}_{ij}\}$ is the $N\times N$
(real-valued) matrix of link ``weights''
\cite{Barr04,Bart05,Barr05,DeMontis2005}. Without loss of
generality, we can assume $\widetilde{a}_{ij}\in [0,1], \forall
i\neq j$ and $\widetilde{a}_{ii}=\widetilde{a} \in\{0,1\},
i,j=1,\dots\,N$ \footnote{We assume that entries in the main
diagonal are either all equal to zero
($\widetilde{a}=0,i=1,\dots\,N$, i.e. no self-interactions) or all
equal to one ($\widetilde{a}=1,i=1,\dots\,N$, i.e.
self-interactions are allowed).}. In line with social network
analysis, we interpret the generic out-of-diagonal entry
$\widetilde{a}_{ij},i\neq j$, as the weight associated to the
directed link originating from node $i$ and ending up in node $j$
(i.e., the strength of the directed link $i\rightarrow j$ in the
graph). A directed edge from $i$ to $j$ is present if and only if
$\widetilde{a}_{ij}>0$.

The idea underlying the construction of the index is very simple.
If the graph $\widetilde{G}$ is undirected, then
$\widetilde{A}=\widetilde{A}^{T}$, where $\widetilde{A}^{T}$ is
the transpose of $\widetilde{A}$. Denoting by $\|\cdot \|$ any
norm defined on a square-matrix, the extent to which
directionality of links counts in the graph $\widetilde{G}$ can
therefore be measured by some increasing function of
$\|\widetilde{A}-\widetilde{A}^{T}\|$, suitably rescaled by some
increasing function of $\|\widetilde{A} \|$ (and possibly of
$\|\widetilde{A}^{T} \|$).

To build the index we first define, again without loss of
generality:

\begin{equation}
A=\{a_{ij}\}= \widetilde{A}-(1-\widetilde{a})I_N,
\end{equation}
where $I_N$ is the $N\times N$ identity matrix. Accordingly, we
define the graph $G=(N,A)$. Notice that
$a_{ij}=\widetilde{a}_{ij}$ for all $i\neq j$, while now
$a_{ii}=1$ for all $i$.

Consider then the square of the Frobenius (or Hilbert-Schmidt)
norm:

\begin{equation}
\|A\|_F^2=\sum_{i}\sum_{j}{a_{ij}^2}=N+\sum_{i}\sum_{j\neq
i}{a_{ij}^2},
\end{equation}
where all sums (also in what follows) span from 1 to $N$. Notice
that $\|A\|_F^2$ is invariant with respect to the transpose
operator, i.e. $\|A\|_F=\|A^{T}\|_F$.

We thus propose the following index:

\begin{equation}
\widetilde{S}(A)=\frac{\|A-A^{T}\|_F^2}{\|A\|_F^2+\|A^{T}\|_F^2}=\frac{\|A-A^{T}\|_F^2}{2\|A\|_F^2}
=\frac{1}{2}\left[\frac{\|A-A^{T}\|_F}{\|A\|_F}\right]^2.
\label{eq:index_s_tilde}
\end{equation}
By exploiting the symmetry of $(a_{ij}-a_{ji})^2$, one easily
gets:

\begin{equation}
\widetilde{S}(A)=\frac{\sum_{i}\sum_{j}{(a_{ij}-a_{ji})^2}}
{2\sum_{i}\sum_{j}{a_{ij}^2}}=\frac{\sum_{i}\sum_{j>i}{(a_{ij}-a_{ji})^2}}
{N+\sum_{i}\sum_{j\neq i}{a_{ij}^2}}.
\end{equation}
Alternatively, by expanding the squared term at the numerator, we
obtain:

\begin{eqnarray}
\widetilde{S}(A)=1-\frac{\sum_{i}\sum_{j}{a_{ij}a_{ji}}}
{\sum_{i}\sum_{j}{a_{ij}^2}}=1-\frac{N+\sum_{i}\sum_{j\neq
i}{a_{ij}a_{ji}}}{N+\sum_{i}\sum_{j\neq i}{a_{ij}^2}}=\\
=1-\frac{N+2\sum_{i}\sum_{j>i}{a_{ij}a_{ji}}}{N+\sum_{i}\sum_{j\neq
i}{a_{ij}^2}}=\frac{\sum_{i}\sum_{j\neq
i}{a_{ij}^2}-2\sum_{i}\sum_{j>i}{a_{ij}a_{ji}}}{N+\sum_{i}\sum_{j\neq
i}{a_{ij}^2}}.\label{eq:max}
\end{eqnarray}
The index $\widetilde{S}(A)$ has a few interesting properties,
which we summarize in the following:

\begin{lemma}[General properties of $\widetilde{S}$]\label{lemma:properties}
For all real-valued $N\times N$ matrices $A=\{a_{ij}\}$ s.t.
$a_{ij}\in [0,1], i\neq j$ and $a_{ii}=1, i=1,...,N$, then:
\begin{enumerate} \item[(1)] $\widetilde{S}(A)\geq 0$. \item[(2)]
$\widetilde{S}(A)=0 \Leftrightarrow A=A^T$, i.e. if and only if
the graph is undirected. \item[(3)] $\widetilde{S}(A)\leq
\frac{N-1}{N+1}$
\end{enumerate}
\end{lemma}
\begin{proof}
See Appendix \ref{Sec:Appendix_1}.
\end{proof}
\bigskip

Furthermore, when $G$ is binary (i.e., $a_{ij}\in \{0,1\}$ for all
$i,j$), the index in eq. \ref{eq:index_s_tilde} turns out to be
closely connected to the density of the graph (i.e., the ratio
between the total number of directed links to the maximum possible
number of directed links) and the ratio of the number of bilateral
directed links in $G$  (i.e. links from $i$ to $j$ s.t.
$a_{ij}=a_{ji}=1$) to the maximum possible number of directed
links. More precisely:

\begin{lemma}[Properties of $\widetilde{S}$ in the case of binary
graphs]\label{lemma:properties_binary} When $G$ is binary, i.e.
$a_{ij}\in \{0,1\}$, all $i,j$, then:
\begin{equation}
\widetilde{S}(A)=\frac{d(A)-b(A)}{(N-1)^{-1}+d(A)}.
\end{equation}
where $d(A)$ is the density of $G$ and $b(A)$ is the ratio between
the number of bilateral directed links to the maximum number of
directed links.
\end{lemma}
\begin{proof}
See Appendix \ref{Sec:Appendix_2}.
\end{proof}

\bigskip
Notice that, in the case of undirected graphs, $b(A)=d(A)$ and
$\widetilde{S}(A)=0$. On the contrary, when there are no bilateral
links, $b(A)=0$. Hence, $\widetilde{S}(A)=[d(A)/(N-1)+1]^{-1}$,
which is maximized when $d(A)=\frac{1}{2}$, i.e.
$\widetilde{S}(A)=\frac{N-1}{N+1}$, as shown in Lemma
\ref{lemma:properties}. Obviously, the larger $b(A)$, the more the
graph $G$ is undirected. As mentioned in Section
\ref{Section:Introduction}, $b(A)$ can be employed to check for
the extent to which directionality counts in $G$. However, such
index is not very useful in weighted graphs, as it does not take
into account the size effect (i.e. the size of weights as measured
by $a_{ij}\in[0,1]$).

In the case of binary graphs, the index in
\eqref{eq:index_s_tilde} almost coincides with the one proposed in
\cite{GarlaSymm2004}. They suggest to employ the correlation
coefficient between $(a_{ij},a_{ji})$ entries in the adjacency
matrix (excluding self loops). It can be shown that this
alternative index -- unlike the one in \eqref{eq:index_s_tilde}
 -- increases with the number of reciprocated links, only if both the
total number of links that are in place and the density of the
network remain constant. See also \cite{Garla2004,Garla2005} for
applications.

Since $\widetilde{S}(A)\in[0,\frac{N-1}{N+1}]$, in what follows we
shall employ its rescaled version:

\begin{equation}
S(A)=\frac{N+1}{N-1}\widetilde{S}(A), \label{eq:index_s}
\end{equation}
which ranges in the unit interval and thus has a more
straightforward interpretation.

\section{Statistical Properties\label{Section:Statistical_Properties}}
\noindent In this section we study the distribution of the index
$S$ as defined in eqs. \ref{eq:index_s_tilde} and
\ref{eq:index_s}. Indeed, despite the range of $S$ does not depend
on $N$, we expect its distribution to be affected by: (i) the size
of the matrix ($N$); (ii) whether the underlying graph $G$ is
binary ($a_{ij}\in \{0,1\}$) or weighted ($a_{ij}\in [0,1]$).

To do so, for each $N\in\{5,10,50,100,200,500,700,1000\}$ we
generate $M=100,000$ random matrices $A$ obeying the restriction
that $a_{ii}=1$, all $i$. In the binary case, out-of-diagonal
entries $\{a_{ij},i\neq j\}$ are drawn from i.i.d. Bernoulli
random variables with $prob\{a_{ij}=0\}=prob\{a_{ij}=1\}=0.5$. In
the weighted case, entries $a_{ij}$ are i.i.d. random variables
uniformly-distributed over $[0,1]$. We then estimate the
distributions of $S$ in both the binary and the weighted cases,
and we study their behavior as the size of the graph increases.
Let us denote by $m_B(N)$ (respectively, $m_W(N)$) the sample mean
of the index $S$ in the binary (respectively, weighted) case, and
by $s_B(N)$ (respectively, $s_W(N)$) the sample standard deviation
of the index $S$ in the binary (respectively, weighted) case.
Simulation results are summarized in the following points.

\begin{enumerate}

        \item In both the binary and the weighted case, the index $S$
        approximately distributes as a Beta random variable for all $N$.
        As $N$ increases, $m_B(N)$ decreases towards 0.50 whereas $m_W(N)$
        increases towards 0.25. Both standard deviations decrease towards
        0. More precisely, the following approximate relations
        hold (see Figures \ref{Fig:scaling_bern} and \ref{Fig:scaling_unif}):

        \begin{eqnarray}
        m_{B}(N)\simeq 0.50+exp\{-1.786369-1.680938lnN\}\label{eq:mean_binary}\\
        m_{W}(N)\simeq 0.25-exp\{-1.767551-0.937586lnN\}\label{eq:mean_weighted}\\
        s_{B}(N)\simeq exp\{-0.135458-1.001695lnN\}\label{eq:sd_binary}\\
        s_{W}(N)\simeq exp\{-0.913297-0.982570lnN\}\label{eq:sd_weighted}
        \end{eqnarray}

        \begin{figure}[h]
        \begin{minipage}[h]{7.5cm}
        \begin{scriptsize}
        \centering {\includegraphics[width=7.5cm]{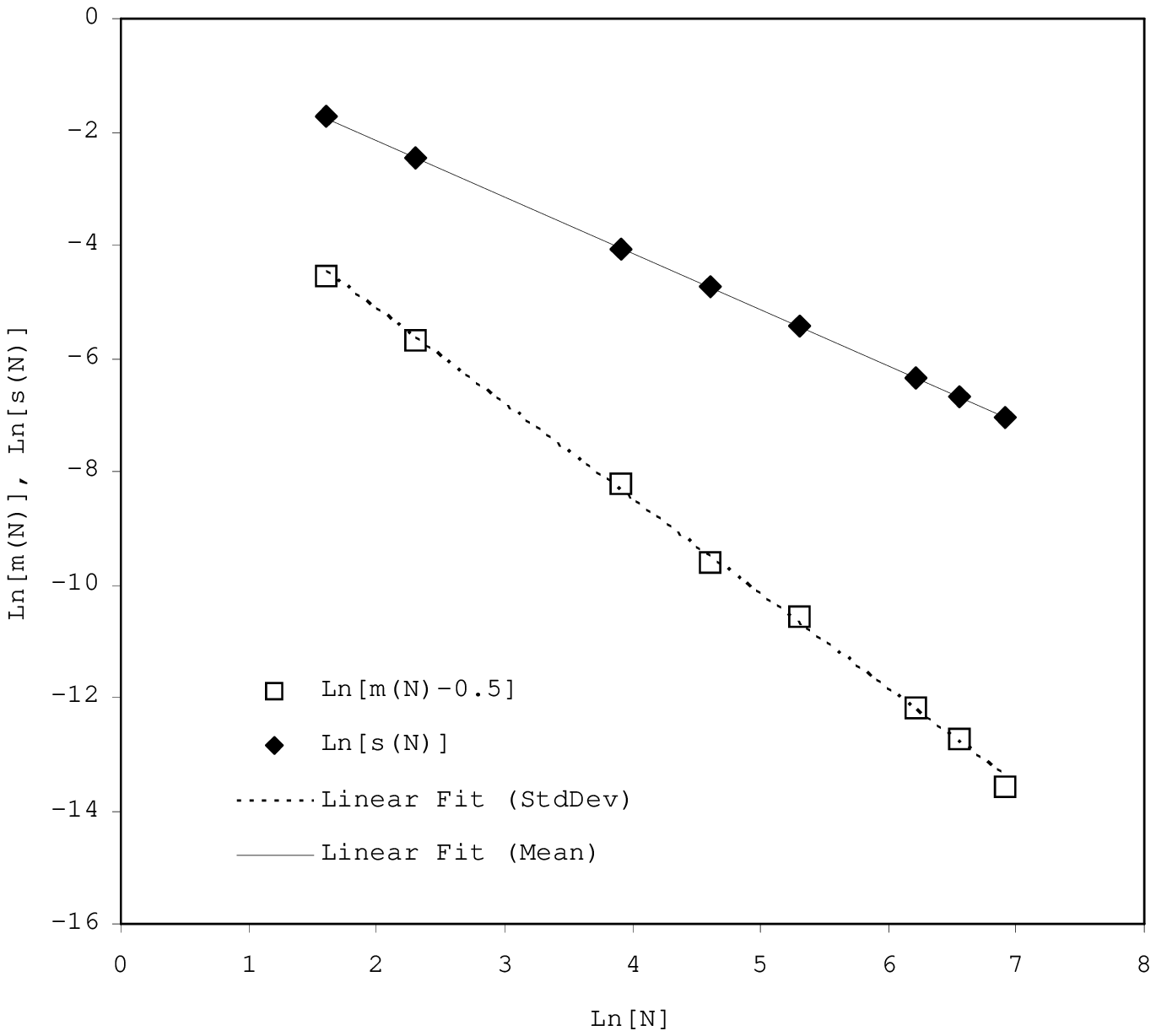}}
        \caption{Binary Graphs. Sample mean and standard deviation of $S$ vs. $N$,
        together with OLS fits. Log-scale on both axes. OLS fits:
        $ln[m_{B}(N)-0.50]\simeq -1.786369-1.680938lnN$ ($R^2=0.998934$)
        and $ln[s_{B}(N)]\simeq -0.135458-1.001695lnN$ ($R^2=0.999995$).}
        \label{Fig:scaling_bern}
        \end{scriptsize}
        \end{minipage}\hfill
        \begin{minipage}[h]{7.5cm}
        \begin{scriptsize}
        \centering {\includegraphics[width=7.5cm]{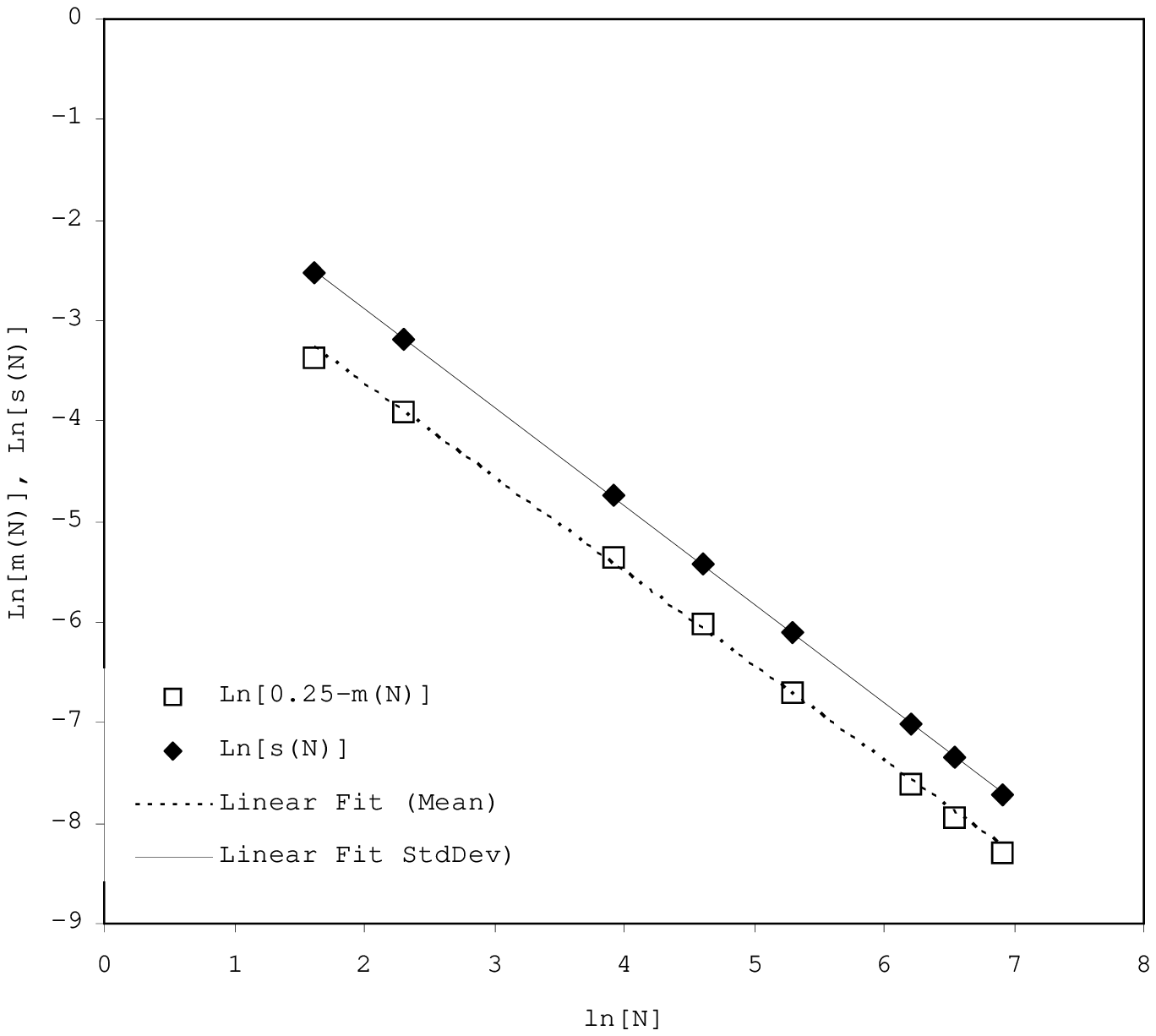}}
        \caption{Weighted Graphs. Sample mean and standard deviation of $S$ vs. $N$,
        together with OLS fits. Log-scale on both axes. OLS fits:
        $ln[0.25-m_{W}(N)]\simeq -1.767551-0.937586lnN$ ($R^2=0.998966$)
        and $ln[s_{W}(N)]\simeq -0.913297-0.982570lnN$ ($R^2=0.999932$).}
        \label{Fig:scaling_unif}
        \end{scriptsize}
        \end{minipage}
        \end{figure}

        \item Given the approximate relations
        in eqs. \ref{eq:mean_binary}-\ref{eq:sd_weighted}, let us standardize
        the index $S$ as follows:

        \begin{eqnarray}
        S_{B}(A)=\frac{S(A)-m_{B}(N)}{s_{B}(N)},\\
        S_{W}(A)=\frac{S(A)-m_{W}(N)}{s_{W}(N)}.
        \end{eqnarray}

        Simulations indicate that the standardized versions of the
        index, i.e. $S_{B}$ and $S_{W}$, are both well approximated by a
        $N(0,1)$, even for small $N$s ($N\geq 10$). Indeed, as
        Figures \ref{Fig:rescaled_bern} and \ref{Fig:rescaled_unif}
        show, the mean of the distributions of $S_{B}$
        and $S_{W}$ vs. $N$ converges towards zero, while the standard
        deviation approaches one (we actually plot standard deviation
        minus one to have a plot in the same scale). Also the third
        (skewness) and the fourth moment (excess kurtosis) stay close
        to zero. We also plot the estimated distribution of $S_{B}$
        and $S_{W}$ vs. $N$, see Figures \ref{Fig:distr_bern} and
        \ref{Fig:distr_unif}. It can be seen that all estimated densities
        collapse towards a $N(0,1)$. Notice that the y-axis is in log scale:
        this allows one to appreciate how close to a $N(0,1)$ are the distributions
        for all $N$ on the tails.

        \begin{figure}[h]
        \begin{minipage}[h]{7.5cm}
        \begin{scriptsize}
        \centering {\includegraphics[width=7.5cm]{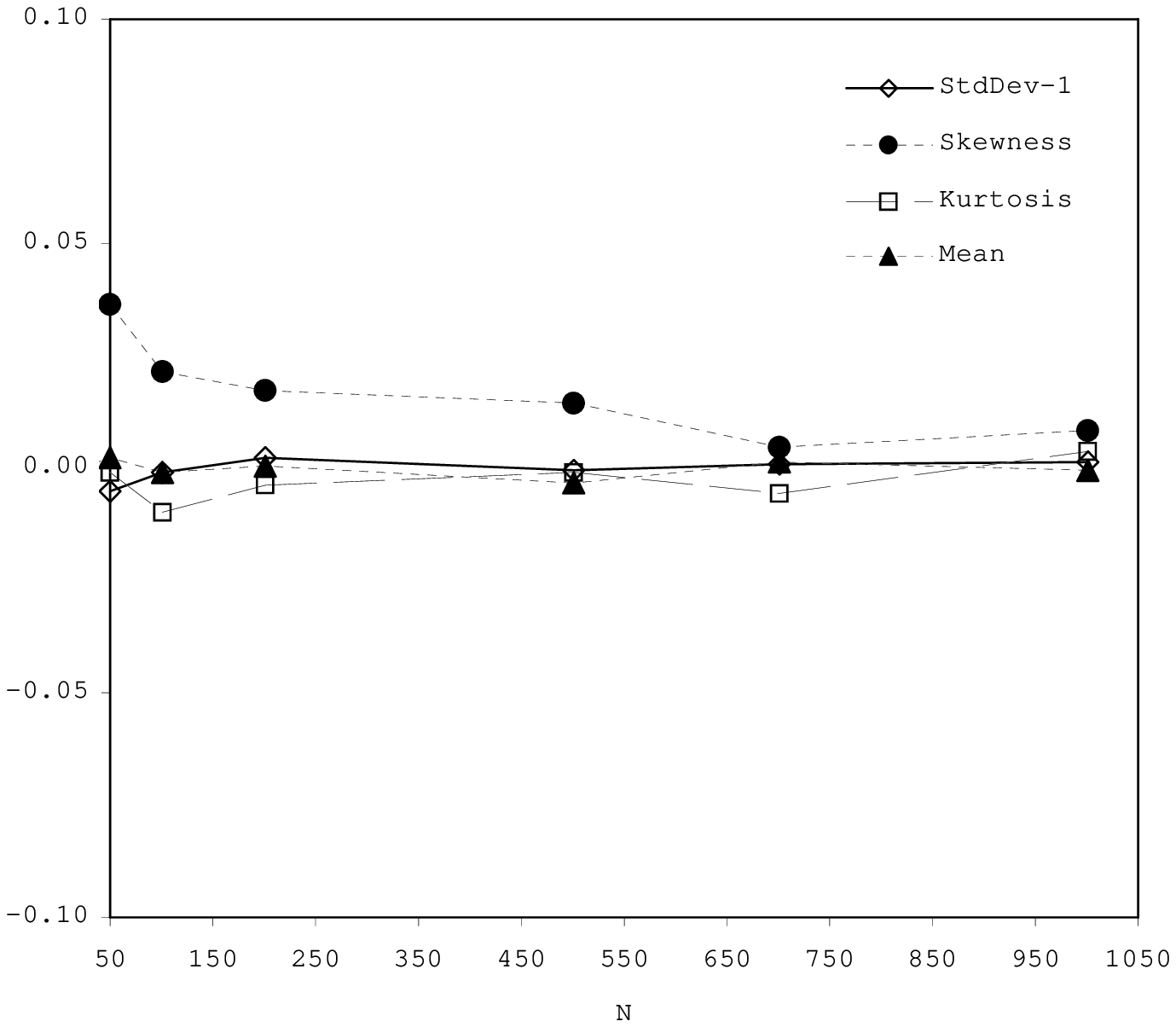}}
        \caption{Binary Graphs. Moments of  $S_{B}$ vs. $N$.} \label{Fig:rescaled_bern}
        \end{scriptsize}
        \end{minipage}\hfill
        \begin{minipage}[h]{7.5cm}
        \begin{scriptsize}
        \centering {\includegraphics[width=7.5cm]{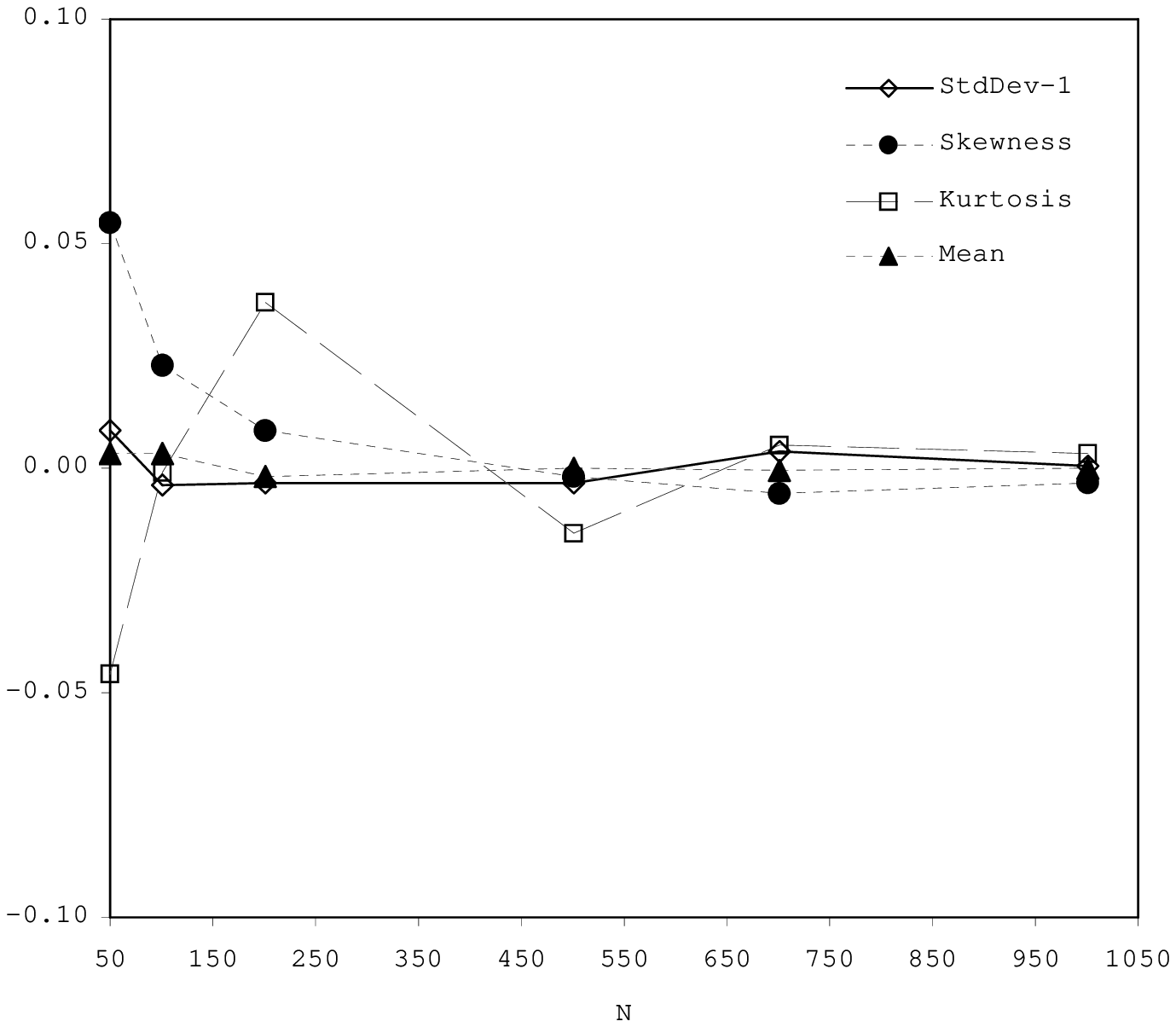}}
        \caption{Weighted Graphs. Moments of  $S_{W}$ vs. $N$.} \label{Fig:rescaled_unif}
        \end{scriptsize}
        \end{minipage}
        \end{figure}

        \begin{figure}[h]
        \begin{minipage}[h]{7.5cm}
        \begin{scriptsize}
        \centering {\includegraphics[width=7.5cm]{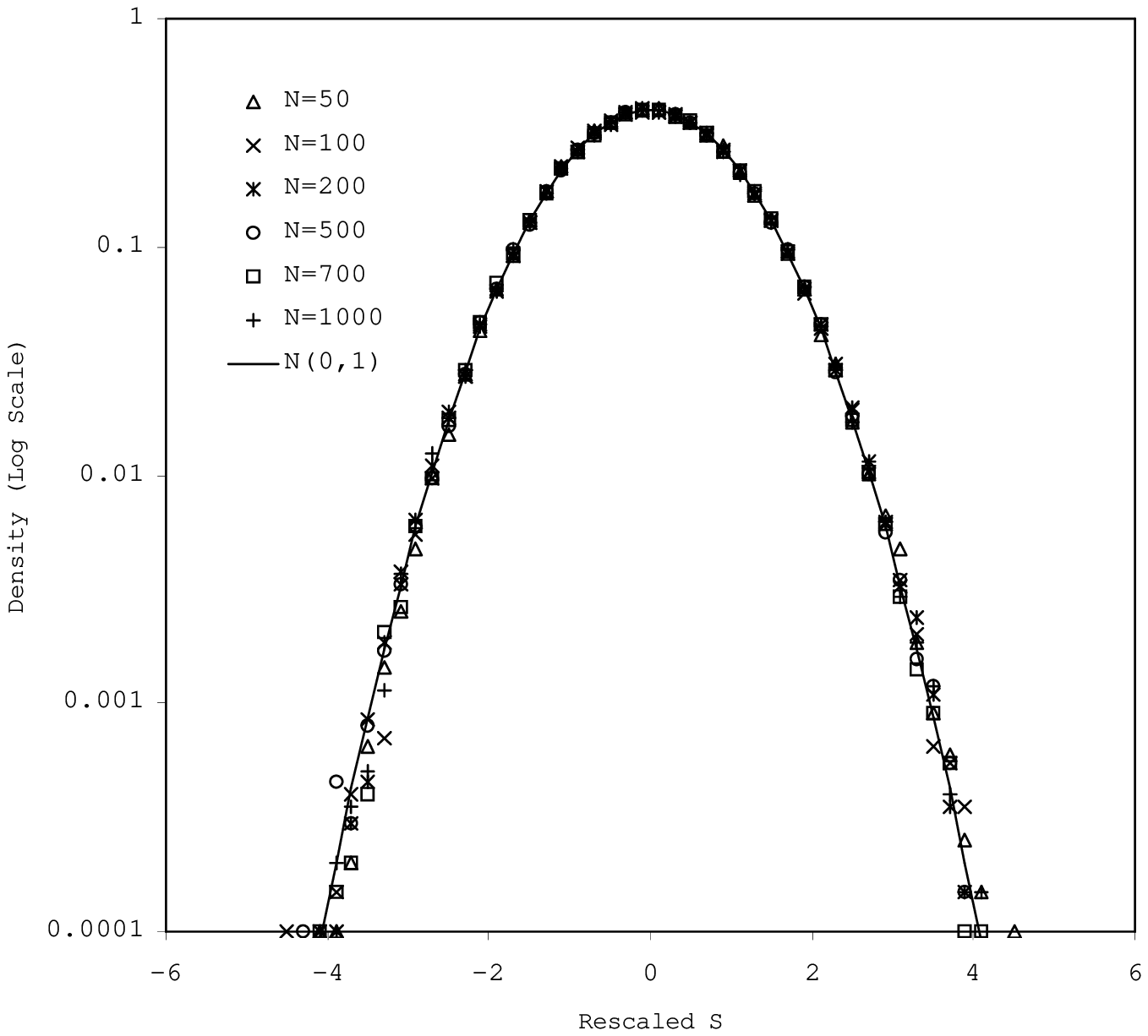}}
        \caption{Binary Graphs. Estimated distribution of $S_{B}$ vs. $N$.
        The $N(0,1)$ fit is also shown as a solid line.}\label{Fig:distr_bern}
        \end{scriptsize}
        \end{minipage}\hfill
        \begin{minipage}[h]{7.5cm}
        \begin{scriptsize}
        \centering {\includegraphics[width=7.5cm]{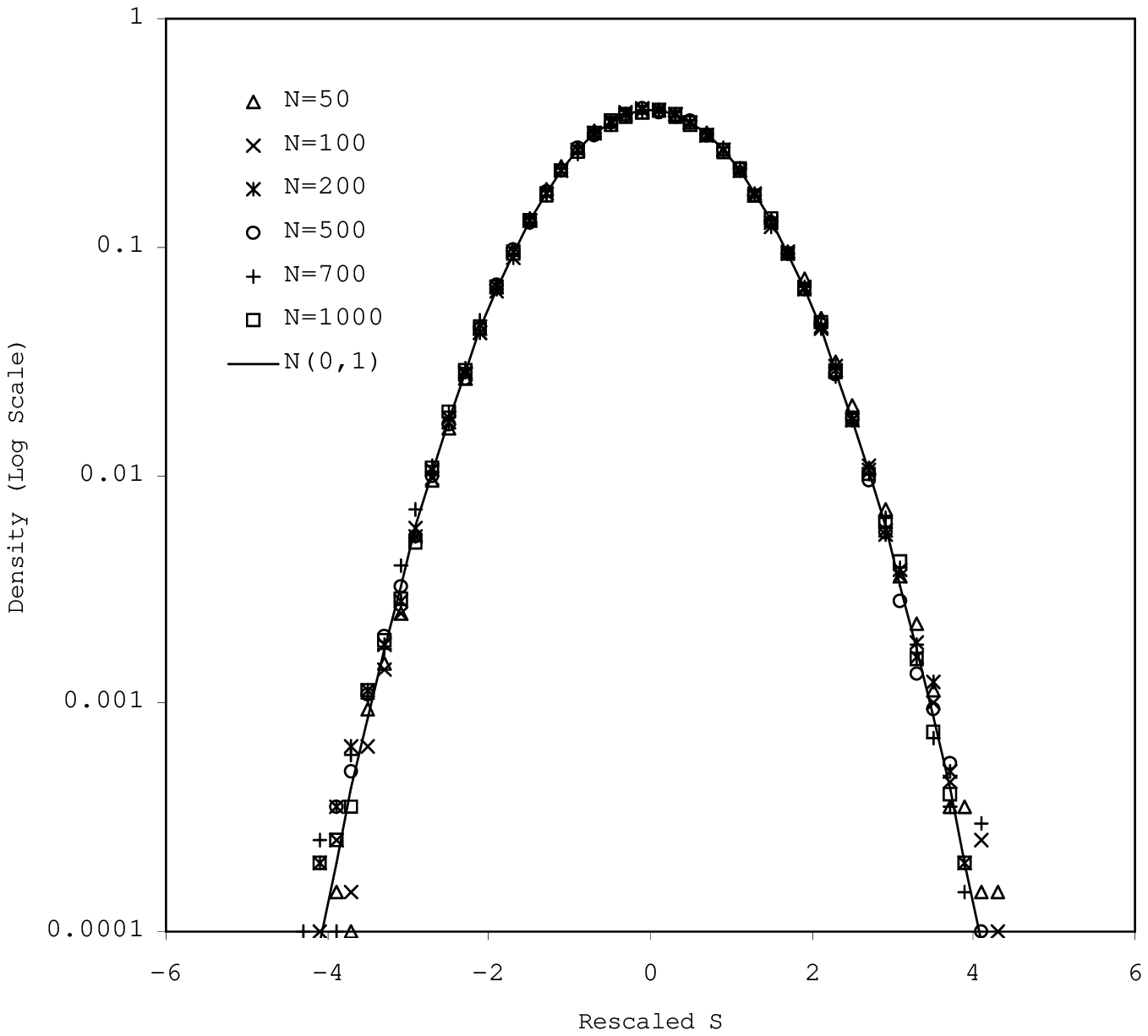}}
        \caption{Weighted Graphs. Estimated distribution of $S_{W}$ vs. $N$.
        The $N(0,1)$ fit is also shown as a solid line.}\label{Fig:distr_unif}
        \end{scriptsize}
        \end{minipage}
        \end{figure}

        Notice finally that as $N$ increases, the distribution maintains a
        constant second moment but the range increases linearly
        with $N$, see \ref{Fig:bounds_bern} and \ref{Fig:bounds_unif}.
        The lower bound (LB) and the upper bound (UB) indeed read approximately:

        \begin{figure}[h]
        \begin{minipage}[h]{7.5cm}
        \begin{scriptsize}
        \centering {\includegraphics[width=7.5cm]{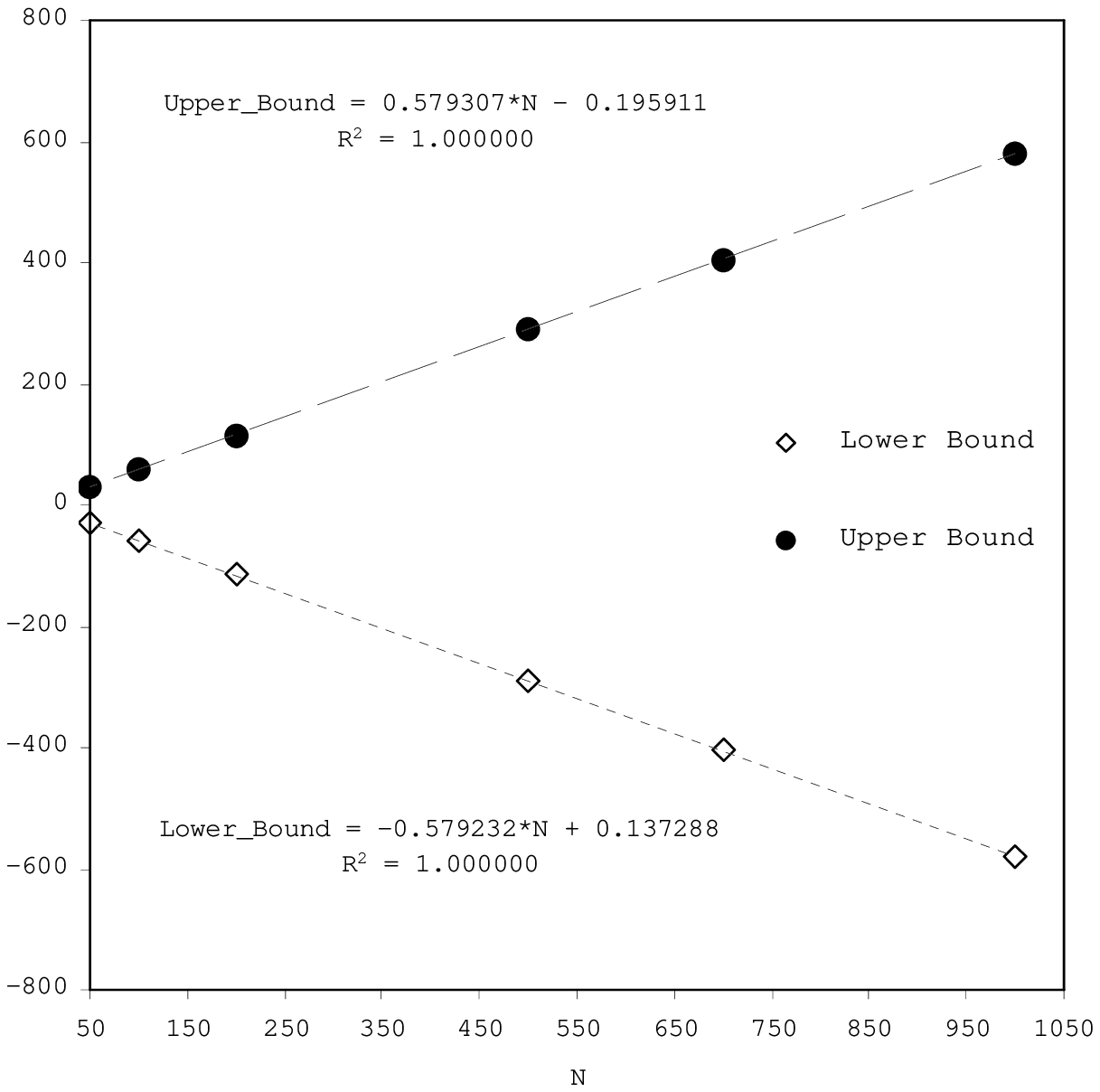}}
        \caption{Binary Graphs. Lower and upper bounds of the re-scaled index $S_{B}$ vs.
        $N$, together with the OLS fit.}\label{Fig:bounds_bern}
        \end{scriptsize}
        \end{minipage}\hfill
        \begin{minipage}[h]{7.5cm}
        \begin{scriptsize}
        \centering {\includegraphics[width=7.5cm]{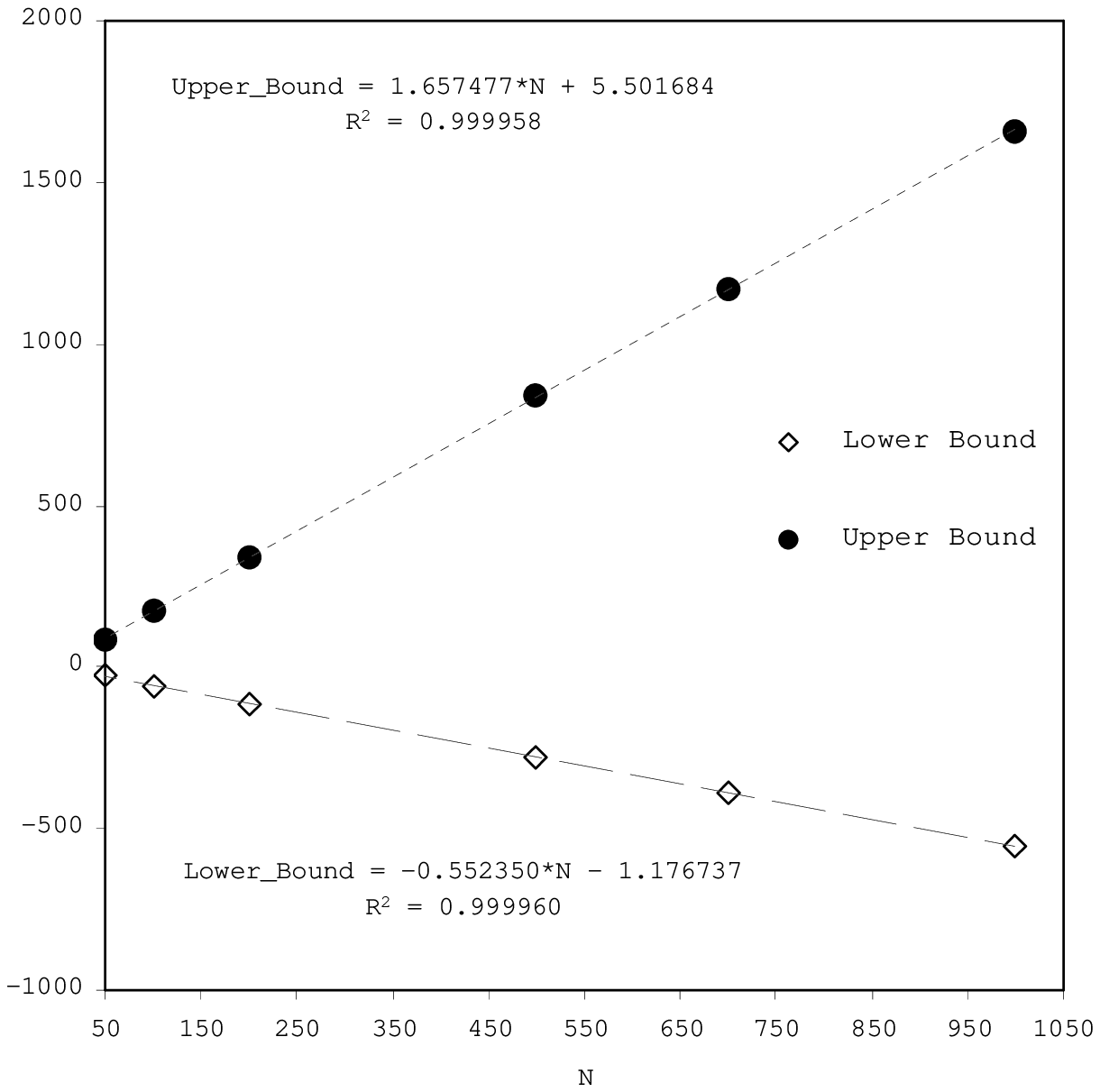}}
        \caption{Weighted Graphs. Lower and upper bounds of the re-scaled index $S_{W}$ vs.
        $N$, together with the OLS fit.}\label{Fig:bounds_unif}
        \end{scriptsize}
        \end{minipage}
        \end{figure}

        \begin{equation} \label{eq:bounds}
        LB_{\ast}(N)=\simeq -\frac{m_{\ast}(N)}{s_{\ast}(N)},
        UB_{\ast}(N)=\simeq \frac{1-m_{\ast}(N)}{s_{\ast}(N)}
        \end{equation}
        where $\{\ast\}=\{B,W\}$ stands for binary (B) and weighted (W).
        Since the standardized index is well approximated
        by a $N(0,1)$ for all $N$, this means that extreme values become
        more and more unlikely. This is intuitive, because as $N$ grows the number of
        matrices with highest/lowest values of the index are very rare.

\end{enumerate}

\section{Examples\label{Section:Examples}}

\noindent The index developed above can be easily employed to
assess the extent to which link directionality matters in
real-world networks. Let us suppose to have estimated a $N\times
N$ matrix $X=\{x_{ij}\}$ describing a binary (B) or a weighted (W)
graph. We then compute the index:

\begin{eqnarray}
S_{\ast}(X)=\frac{\frac{N+1}{N-1}\widetilde{S}(X)-m_{\ast}(N)}{s_{\ast}(N)}=\\
=\frac{1}{s_{\ast}(N)}\left[\frac{N+1}{N-1}\frac{\sum_{i}\sum_{j>i}{(x_{ij}-x_{ji})^2}}{N+\sum_{i}\sum_{j\neq
i}{x_{ij}^2}}-m_{\ast}(N)\right].
\end{eqnarray}
where $\{\ast\}=\{B,W\}$ and $(m_{\ast}(N),s_{\ast}(N))$ are as in
eqs. \ref{eq:mean_binary}-\ref{eq:sd_weighted}. Since we know that
$S_{\ast}(X)$ is approximately $N(0,1)$, we can fix a lower
threshold in order to decide whether the network is sufficiently
(un)directed. For instance, we could set the lower threshold equal
to 0 (i.e. equal to the mean), and decide that if $S_{\ast}(X)>0$
(above the mean) we shall treat the network as directed (and
undirected otherwise). More generally, one might set a threshold
equal to $x\in R$ and conclude that the graph is undirected if
$S_{\ast}<x$. On the contrary, one should expect the directional
nature of the graph to be sufficiently strong, so that a digraph
analysis is called for.

To test the index against real-world cases, we have taken the
thirteen social and economic networks analyzed in
\cite{WassermanFaust1994}, see Table \ref{Tab:Example}
\footnote{They concern advice, friendship and ``reports-to''
relations among Krackhardt's high-tech managers; business and
marital relationships between Padgett's Florentine families;
acquaintanceship among Freeman's EIES researchers and messages
sent between them; and data about trade flows among countries (cf.
\cite{WassermanFaust1994}, ch. 2.5 for a thorough description).}.
All networks are binary and directed, apart from Freeman's ones
(which are \textit{weighted} and directed) and Padgett's ones
(which are binary and \textit{undirected}). Table
\ref{Tab:Example} reports both the index $S$ and its standardized
versions $S_{\ast}, \{\ast\}=\{B,W\}$, for all cited examples.

\begin{table}[h]
\caption{The index $S$ and its standardized version $S_{\{\ast\}},
\{\ast \}=\{B(inary),W(eigthed)\}$ for social networks studied in
\cite{WassermanFaust1994}, cf. Chapter 2.5.} \label{Tab:Example}
\begin{ruledtabular}
\begin{tabular}{rlrrr}
& Social Network & $N$ & $S$ & $S_{\ast}$\\
\hline
1 & Advice relations btw Krackhardt's hi-tech managers &     21 &     0.521327 & 0.491228    \\
2 & Friendship relations btw Krackhardt's hi-tech managers &     21 &     0.500813 & 0.004610   \\
3 & ``Reports-to'' relations btw Krackhardt's hi-tech managers &     21 &     0.536585 & 0.860033   \\
\hline
4 & Business relationships btw Padgett's Florentine families &     16 &     0.000000 & -9.232823   \\
5 & Marital relationships btw Padgett's Florentine families &     16 &     0.000000 & -9.232823   \\
\hline
6 & Acquaintanceship among Freeman's EIES researchers (Time 1)&     32 &     0.109849 & -10.025880   \\
7 & Acquaintanceship among Freeman's EIES researchers (Time 2) &     32 &     0.094968 & -11.143250   \\
8 & Messages sent among Freeman's EIES researchers &     32 &     0.014548 & -17.181580   \\
\hline
9 & Country Trade Flows: Basic Manufactured Goods&     24 &     0.260349 & -6.643695   \\
10 & Country Trade Flows: Food and Live Animals&     24 &     0.311966 & -5.217508   \\
11 & Country Trade Flows: Crude Materials (excl. Food)&     24 &     0.272560 & -6.306300   \\
12 & Country Trade Flows: Minerals, Fuels, Petroleum &     24 &     0.403336 &  -2.692973  \\
13 & Country Trade Flows: Exchange of Diplomats&     24 &     0.080208 & -11.620970   \\
\end{tabular}
\end{ruledtabular}
\end{table}

Suppose to fix the lower threshold equal to zero. Padgett's
networks, being undirected, display a very low value (in fact, the
non standardized index is equal to zero as expected). The table
also suggests to treat all the binary trade networks as
undirected. The same advice applies for Freeman's networks, which
are instead weighted. The only networks which have an almost clear
directed nature (according to our threshold) are Krackhardt's
ones. In that case our index indicates that a directed graph
analysis would be more appropriate.

\section{Concluding Remarks \label{Section:Conclusions}}

\noindent In this paper we have proposed a new procedure that
might help to decide whether an empirically-observed adjacency or
weights $N\times N$ matrix $W$, describing the graph underlying a
social or economic network, is sufficiently symmetric to justify
an undirected network analysis. The index that we have developed
has two main properties. First, it can be applied to both binary
or weighted graphs. Second, once suitably standardized, it
distributes as a standard normal over all possible
adjacency/weights matrices. Therefore, given a threshold decided
by the researcher, any empirically observed adjacency/weights
matrix displaying a value of the index lower (respectively,
higher) than the threshold is to be treated as if it characterizes
an undirected (respectively, directed) network.

It must be noticed that setting the threshold always relies on a
personal choice, as also happens in statistical hypothesis tests
with the choice of the significance level $\alpha$. Despite this
unavoidable degree of freedom, the procedure proposed above still
allows for a sufficient comparability among results coming from
different studies (i.e. where researchers set different threshold)
if both the value of the index $S$ and the size of the network are
documented in the analysis. In that case, one can easily compute
the probability of finding a matrix with a lower/higher degree of
symmetry, simply by using the definition of bounds (see eq.
\ref{eq:bounds}) and probability tables for the standard normal.

A final remark is in order. As mentioned, our procedure does not
configure itself as a statistical test. Since the researcher often
relies on a single observation of the network under study (or a
sequence of serially-correlated network snapshots through time),
statistical hypothesis testing will be only very rarely feasible.
Nevertheless, in the case where a sample of $M$ i.i.d.
observations of $W$ is available, one might consider to use the
the sample average of the index (multiplied by $\sqrt{M}$) and
employ the cental limit theorem to test the hypothesis that the
observations come from a undirected (random) graph.

\begin{acknowledgments}
\noindent This work has enormously benefited from insightful
suggestions by Michel Petitjean. Thanks also to Javier Reyes and
Stefano Schiavo for their useful comments.
\end{acknowledgments}

\bibliographystyle{apsrev}
%\bibliography{symmetry}

\begin{appendix}

\section{Proof of Lemma \ref{lemma:properties}}\label{Sec:Appendix_1}
\noindent Points (1) and (2) simply follow from the definition in
eq. \ref{eq:index_s_tilde}. As to (3), let us suppose that there
exists a matrix $A$ satisfying the above restrictions and such
that $\widetilde{S}(A)>\frac{N-1}{N+1}$. Then, using eq.
\ref{eq:max}:

\begin{equation}
\frac{N+2\sum_{i}\sum_{j>i}{a_{ij}a_{ji}}}{N+\sum_{i}\sum_{j\neq
i}{a_{ij}^2}}<\frac{2}{N+1}.
\end{equation}
The best case for such an inequality to be satisfied is when the
the left hand side is minimized. This is achieved when there are
$N(N-1)/2$ entries equal to one and $N(N-1)/2$ entries equal to
zero in such a way that $a_{ij}\neq a_{ji}$ for all $i\neq j$
(e.g., when the upper diagonal matrix is made of all ones and the
lower diagonal matrix is made of all zeroes). In that case the
left hand side is exactly equal to $\frac{2}{N+1}$, leading to the
absurd conclusion that $\frac{2}{N+1}<\frac{2}{N+1}$.

\section{Proof of Lemma \ref{lemma:properties_binary}}\label{Sec:Appendix_2}
\noindent It follows from the definition of $d(A)$ that:

\begin{equation}
d(A)=\frac{\sum_{i}\sum_{j\neq
i}{a_{ij}}}{N(N-1)}=\frac{\sum_{i}\sum_{j\neq
i}{a_{ij}^2}}{N(N-1)},
\end{equation}

Moreover, it is easy to see that:

\begin{equation}
b(A)=\frac{\sum_{i}\sum_{j\neq
i}{a_{ij}a_{ji}}}{N(N-1)}=\frac{2\sum_{i}\sum_{j>i}{a_{ij}a_{ji}}}{N(N-1)}.
\end{equation}

To prove the Lemma, it suffices to note that:
\begin{eqnarray}
\widetilde{S}(A)=1-\frac{N+2\sum_{i}\sum_{j>i}{a_{ij}a_{ji}}}{N+\sum_{i}\sum_{j\neq
i}{a_{ij}^2}}=\\
=1-\frac{1+(N-1)b(A)}{1+(N-1)d(A)}=\frac{d(A)-b(A)}{(N-1)^{-1}+d(A)}.
\end{eqnarray}

\end{appendix}
\end{document}